\documentclass[12pt]{article}

\usepackage{authblk}
\usepackage{natbib}
\usepackage{graphicx}%
\usepackage{multirow}%
\usepackage{amsmath,amssymb,amsfonts}%
\usepackage{amsthm}%
\usepackage{mathrsfs}%
\usepackage[title]{appendix}%
\usepackage{textcomp}%
\usepackage{manyfoot}%
\usepackage{booktabs}%
\usepackage{algorithm}%
\usepackage{algorithmicx}%
\usepackage{algpseudocode}%
\usepackage{listings}%
\usepackage{orcidlink}
\usepackage{setspace}
\usepackage[labelsep=space]{caption}
\usepackage{dsfont}
\usepackage{bbold}
\usepackage{breqn}
\usepackage{multicol}
\usepackage{float}
\usepackage{subcaption}
\DeclareMathAlphabet\mathbfcal{OMS}{cmsy}{b}{n}

\DeclareMathOperator*{\argmin}{arg\,min}

\def\boxit#1{\vbox{\hrule\hbox{\vrule\kern6pt
          \vbox{\kern6pt#1\kern6pt}\kern6pt\vrule}\hrule}}

\begin{document}

\def\spacingset#1{\renewcommand{\baselinestretch}%
{#1}\small\normalsize} \spacingset{1}

\title{\bf Air-HOLP: Adaptive Regularized Feature Screening for High Dimensional Correlated Data}
\author[1]{Ibrahim Joudah}
\author[1,2]{Samuel Muller}
\author[1]{Houying Zhu}
\affil[1]{School of Mathematical and Physical Sciences, Macquarie University}
\affil[2]{School of Mathematics and Statistics, University of Sydney}
\maketitle

\bigskip
\abstract{Handling high-dimensional datasets presents substantial computational challenges, particularly when the number of features far exceeds the number of observations and when features are highly correlated. A modern approach to mitigate these issues is feature screening. In this work, the High-dimensional Ordinary Least-squares Projection (HOLP) feature screening method is advanced by employing adaptive ridge regularization. The impact of the ridge tuning parameter on the Ridge-HOLP method is examined and Adaptive iterative ridge-HOLP (Air-HOLP) is proposed, a data-adaptive advance to Ridge-HOLP where the ridge-regularization tuning parameter is selected iteratively and optimally for better feature screening performance. The proposed method addresses the challenges of tuning parameter selection in high dimensions by offering a computationally efficient and stable alternative to traditional methods like bootstrapping and cross-validation. Air-HOLP is evaluated using simulated data and a prostate cancer genetic dataset. The empirical results demonstrate that Air-HOLP has improved performance over a large range of simulation settings. We provide R codes implementing the Air-HOLP feature screening method and integrating it into existing feature screening methods that utilize the HOLP formula.}

\vfill

\newpage
\spacingset{1.125}

\section{Introduction}
\label{sec:intro}

Modern advancements in technology have enabled the collection and storage of high-dimensional datasets containing thousands of features across diverse fields such as in machine learning, tomography, tumor classification, social science, and finance \citep{zhu2011model, liu2022model}. Coping with such dimensions presents computational challenges, even for basic ordinary least square regression. Moreover, handling highly correlated features further complicates the challenge. This paper tackles the challenges associated with such datasets, specifically focusing on feature screening in high dimensional correlated settings where the number of features $p$ exceeds the sample size $n$.

We begin by defining the problem setting and notation used throughout. We consider the linear regression model
\begin{equation} \label{eq:linear}
    \boldsymbol{y} = \boldsymbol{X}\boldsymbol{\beta} + \mathbfcal{E},
\end{equation}
where $\boldsymbol{y}$ is the $n \times 1$ response vector, $\boldsymbol{X}$ is the observed $n \times p$ design matrix, $\boldsymbol{\beta}$ is the true but unknown $p \times 1$ vector of regression coefficients, and $\mathbfcal{E}$ is an $n \times 1$ vector of errors. We assume that $\boldsymbol{X}$ collates $n$ realisations of the random $p \times 1$ predictor vector $\boldsymbol{x} = (x_1, \ldots, x_p)^\top$, and $\boldsymbol{y}$ collates $n$ realisations  of the random response $y$ given by $y = \boldsymbol{x}^\top \boldsymbol{\beta} + \epsilon$, where $\epsilon$ is the error with expected value of zero. The features with non-zero corresponding true regression coefficients in Equation \eqref{eq:linear} are referred to as true features.

Feature screening is a dimensionality reduction technique that aims to efficiently eliminate numerous non-true features while retaining all true features. Feature screening is often used prior to model selection to handle the curse of dimensionality \citep{liu2020variable, liu2015selective, fan2009ultrahigh}. In this paper, we build on the existing High-dimensional Ordinary Least-squares Projection (HOLP) method as introduced in \citet{wang2016high}, by incorporating adaptive ridge regularization. We analyze the impact of the ridge tuning parameter on Ridge-HOLP \citep{wang2016high} and propose the Adaptive iterative ridge-HOLP (Air-HOLP), a method that utilizes a data-adaptive approach for selecting the ridge tuning parameter, which shows remarkable gains in feature screening performance. Additionally, we provide R codes implementing the Air-HOLP feature screening method and integrating it into the group HOLP \citep{qiu2020grouped} method which utilizes the HOLP formula. The codes are made available on GitHub at \url{https://github.com/Logic314/Air-HOLP.git}.

The remainder of the paper is structured as follows: Section \ref{S2} provides background on feature screening and motivates our work. Section \ref{S3} introduces the Air-HOLP method. Section \ref{S4} evaluates the performance of Air-HOLP and of two existing methods using simulated data, while Section \ref{S5} applies Air-HOLP to a prostate cancer dataset. Section \ref{S6} evaluates the speed and computational complexity of Air-HOLP, and Section \ref{S7} concludes the paper.

\section{Background} \label{S2} 

\subsection{Feature Screening} \label{S2S1}

The aim of feature screening is to efficiently eliminate as many non-true features as possible while retaining all true features. The sure screening property is often a desirable property of a feature screening method, i.e.\ the method retains all true features with a probability approaching 1 as $n \rightarrow \infty$, a concept introduce by \citet{fan2008sure}. There is a rich literature on feature screening and we refer to \citet{liu2015selective} and \citet{liu2020variable} for a general review. Here we briefly mention the foundations of feature screening.

A popular feature screening method is Sure Independence Screening (SIS) and it ranks features based on their absolute Pearson correlation with the response. Then, a predetermined threshold is used to screen features, such as screening the top $\lceil n/\log(n) \rceil$ features \citep{fan2008sure}. The SIS method relies on restrictive assumptions to guarantee sure screening, including the true features being independently and linearly related to the response variable, limiting its reliability in practice \citep{zhu2011model}.

Several methods have been proposed to address the linearity limitation of SIS. Examples include non-parametric model-fitting approaches such as spline regression \citep{hall2009using, fan2011nonparametric, fan2014nonparametric} and quantile regression \citep{wu2015conditional,zhong2016regularized,chen2019note}, and robust correlation measures such as Distance Correlation \citep{li2012feature}, Ball Correlation \citep{pan2018generic, zhang2019robust}, and Projection Correlation \citep{liu2022model}. Ultimately, these methods measure the marginal relations between the features and the response. Thus, they do not guarantee sure screening when features are marginally independent and jointly dependent on the response \citep{fan2008sure}. \cite{wang2016high} proposed the HOLP method to resolve this issue, where HOLP measures the joint dependence of features on the response rather than solely learning from the marginal information.

\subsection{HOLP and Ridge-HOLP} \label{S2S2}

The HOLP method uses the Moore-Penrose inverse of $\boldsymbol{X}$ to estimate the vector of regression coefficients in the full regression model, that is
\begin{equation} \label{eq:HOLP}
    \boldsymbol{\hat{\beta}} = \boldsymbol{X}^\top(\boldsymbol{X}\boldsymbol{X}^\top)^{-1}\boldsymbol{y}.
\end{equation} 
Then, the features are ranked by the absolute values of their estimated coefficients denoted by $|\hat\beta_j|$, $j=1,\ldots,p$. Then, a predetermined threshold $m$ is used to screen the top $m$ features.
\cite{wang2016high} showed that Equation \eqref{eq:HOLP} suffers from instability when $\boldsymbol{X}\boldsymbol{X}^\top$ is close to degeneracy
or when $n$ is close to $p$. To address this issue, \cite{wang2016high} proposed Ridge-HOLP:
\begin{equation} \label{eq:Ridge-HOLP}
    \boldsymbol{\hat{\beta}}_{r} = \boldsymbol{X}^\top(\boldsymbol{X}\boldsymbol{X}^\top + r \mathds{I}_n)^{-1}\boldsymbol{y},
\end{equation}
where $r$ is the ridge tuning parameter and $\mathds{I}_n$ denotes the identity matrix.

The Ridge-HOLP formula in Equation \eqref{eq:Ridge-HOLP} is equal to the standard ridge formula but is more efficient when $p > n$ as the computational complexity of Ridge-HOLP is $O(n^3 + n^2p)$, while the standard ridge formula is $O(p^3 + p^2n)$. This is because Ridge-HOLP requires an ($n\times p$) by ($p\times n$) multiplication $O(n^2p)$ and an ($n\times n$) matrix inverse $O(n^3)$, while standard ridge requires a ($p\times n$) by ($n\times p$) multiplication $O(p^2n)$ and a ($p\times p$) matrix inverse $O(p^3)$. When $n$ is close to $p$, \cite{wang2016high} recommend using Ridge-HOLP with a fixed tuning parameter $r = 10$. 

The penalty in the Ridge-HOLP formula in Equation \eqref{eq:Ridge-HOLP} also helps to handle multicollinearity. When features are highly correlated, then estimated regression coefficients have large variances \citep{hoerl1970ridge}, which leads to poor ranking accuracy and thus lower screening performance. The penalty in Ridge-HOLP reduces the estimation sensitivity. However, selecting an appropriate tuning parameter $r$ is crucial to balance the bias-variance trade-off and to maximize the screening performance of Ridge-HOLP. 

Multiple feature screening methods have integrated the HOLP and Ridge-HOLP formulas \eqref{eq:HOLP} and \eqref{eq:Ridge-HOLP} into their screening process. Examples of this include the group HOLP method of \cite{qiu2020grouped}, the Dynamic Tilted Current Correlation method of \cite{zhao2021dynamic}, and the HOLP-DF method of \cite{samat2022holp}. In these methods, the HOLP and Ridge-HOLP formulas are an integral part of the screening process.

\subsection{Challenges of Tuning Parameter Selection in High Dimensions} \label{S2S3}

Several methods exist to choose the ridge tuning parameter. Common approaches include bootstrapping \citep{delaney1986use}, cross-validation \citep{allen1971mean, allen1974relationship} and generalized cross-validation \citep{golub1979generalized}. However, such sub-sampling in the Ridge-HOLP method is computationally expensive due to the repeated computation of the Ridge-HOLP formula. To address this, a more efficient formula called multiridge was proposed by \cite{van2021fast} for applying fast cross-validation in ridge regression. The multiridge formula avoids redundant computations of the ridge formula when applied to multiple values of the tuning parameter $r$, reducing the heavy computations to be repeated only $k$  times in k-fold cross-validation. However, k-fold cross-validation is non-deterministic and can therefore introduce instability in feature screening as different data splits may lead to different screened features. Although repeated cross-validation can mitigate this instability, it comes at the cost of increased computational burden \citep{martinez2011empirical}.

To avoid the computational burden and instability associated with sub-sampling, an alternative approach is to use a closed-form formula for selecting the tuning parameter $r$.
Examples of this alternative approach include \cite{ahmad2006comparative, alkhamisi2007monte, batah2008efficiency, dorugade2010alternative, hoerl1970ridge, hoerl1975ridge, jf1976simulation, kibria2003performance, nomura1988almost}. However, the formulas proposed in these papers require $p$ to be smaller than $n$. \cite{cule2013ridge} addressed this limitation by extending the \cite{hoerl1975ridge} approach to handle $p > n$. Nevertheless, their method involves computing the matrix $\boldsymbol{X}^\top\boldsymbol{X}$ and its eigendecomposition, resulting in a time complexity of $O(p^3 + p^2n)$.

\section{The Air-HOLP method} \label{S3}

Air-HOLP is an adaptive iterative variant of Ridge-HOLP which efficiently selects the tuning parameter $r$. We begin by discussing how to ensure the selected $r$ in Ridge-HOLP satisfies the sure screening property. Theorem $3$ of \cite{wang2016high} and \cite{10.1111/rssb.12427} mentions that for the Ridge-HOLP to achieve sure screening, the tuning parameter $r$ must satisfy $r = o(n^{1-(5/2)\tau-\kappa})$, where  $1-7.5\tau-2\kappa-\nu > 0$, and $\tau, \kappa, \nu>0$. It is worth noting that any $r = cn^a$ where $0 \leq a \leq 0.5$ and $c \geq 0$ satisfies the required condition on $r$. Thus, to ensure that we achieve the sure screening property, we confine the tuning parameter to the range $[0, c\sqrt{n}]$ for some positive constant $c$ and choose $c = 1000$ if not otherwise mentioned. 

We now discuss how to select the tuning parameter. To select an appropriate tuning parameter that balances the trade-off between bias and variance we solve
\begin{equation} \label{eq:mse}
    \hat{r} \; = \; \argmin_{0\leq r\leq c\sqrt{n}}{ \frac{1}{n} \left\| \boldsymbol{y}_{0} - \boldsymbol{\hat{y}}_{r} \right\|^2_2},
\end{equation}
where $\| \cdot \|^2_2$ is the squared L2-norm, $\boldsymbol{y}_{0} = \boldsymbol{X}\boldsymbol{\beta}$ and $\boldsymbol{\hat{y}}_{r} = \boldsymbol{X} \boldsymbol{\hat{\beta}}_{r}$. Equation \eqref{eq:mse} is equivalent to
\begin{equation} \label{eq:Air-HOLP}
    \hat{r} \; = \; \argmin_{0\leq r\leq c\sqrt{n}}{\boldsymbol{\hat{y}}_{r}^\top \boldsymbol{\hat{y}}_{r} - 2 \boldsymbol{y}_{0}^\top \boldsymbol{\hat{y}}_{r}}.
\end{equation}
The choice of $\boldsymbol{y}_0$ in Equation \eqref{eq:mse} is to circumvent the overfitting that occurs when using $\boldsymbol{y}$ instead. If $\boldsymbol{y}$ was used instead of $\boldsymbol{y}_0$, the solution would always be $\hat{r} = 0$ because $\boldsymbol{\hat{y}}_{r} = \boldsymbol{y}$ for $r = 0$ and $p \geq n$. We evade fitting the noise in $\boldsymbol{y}$ by using $\boldsymbol{y}_0$ which represents the noise-free signal.
The challenge in solving Equation \eqref{eq:Air-HOLP} lies in deriving a computationally efficient and accurate estimate of the unknown expected response $\boldsymbol{y}_0$. We address this by proposing the following approach:
\begin{itemize}
    \item First, choose an initial tuning parameter $r_0$ and use it to compute the initial Ridge-HOLP estimator $\boldsymbol{\hat{\beta}}_{r_0} = \boldsymbol{X}^\top(\boldsymbol{X}\boldsymbol{X}^\top + r_0 \mathds{I}_n)^{-1}\boldsymbol{y}$.
    \item Then, rank the features in $\boldsymbol{X}$ by the absolute values of their estimated coefficients $|\boldsymbol{\hat{\beta}}_{r_0}|$ and screen the top $m'$ features, where $m' < n$.
    \item Finally, fit a regression model utilizing the screened features, and compute the fitted response $\boldsymbol{\Tilde{y}}_{r_0}$, which serves as the estimate for $\boldsymbol{y}_0$.
\end{itemize}

Notably, the proposed process does not detail a specific method for fitting the model using the screened features. For simplicity, we employ ordinary least squares regression in our empirical research below. Moreover, we use the initial tuning parameter $r_0 = 10$ following \cite{wang2016high} and choose $m' = \lceil n/\log(n) \rceil$ unless otherwise mentioned. Note that $m'$ needs to be smaller than $n$ to facilitating fitting an ordinary least squares regression model.

Once $\boldsymbol{y}_0$ is estimated, the initial tuning parameter is updated, and the process is repeated iteratively. A given tuning parameter $r_i$ is updated through
\begin{equation} \label{eq:ri}
    r_{i+1} = \argmin_{0\leq r\leq c\sqrt{n}}{\boldsymbol{\hat{y}}_{r}^\top \boldsymbol{\hat{y}}_{r} - 2 \boldsymbol{\Tilde{y}}_{r_i}^\top \boldsymbol{\hat{y}}_{r}}.
\end{equation}
Air-HOLP iteratively updates the tuning parameter $r$ until $|r_{i+1} - r_{i}| < \delta r_{i+1}$ for some small $\delta > 0$, i.e.\ until the absolute relative error is less than $\delta$ or until a predetermined maximum number of iterations is reached.

To solve Equation \eqref{eq:ri} efficiently, we use the eigen decomposition 
\begin{equation} \label{eq:eigen}
    \boldsymbol{X}\boldsymbol{X}^\top = \boldsymbol{U}\boldsymbol{D}_n\boldsymbol{U}^{-1},
\end{equation}
where $\boldsymbol{D}_n$ is a diagonal matrix and $\boldsymbol{U}\boldsymbol{U}^\top = \mathds{I}_n$. The eigen decomposition in Equation \eqref{eq:eigen} allows to compute $(\boldsymbol{X}\boldsymbol{X}^\top + r\mathds{I}_n)^{-1}$ by $\boldsymbol{U}(\boldsymbol{D}_{n} + r \mathds{I}_n)^{-1}\boldsymbol{U}^{\top}$. Thus, $\boldsymbol{\hat{y}}_{r} = \boldsymbol{U}\boldsymbol{D}_{n}(\boldsymbol{D}_{n} + r \mathds{I}_n)^{-1}\boldsymbol{U}^{\top}\boldsymbol{y}$. Substituting the eigen decomposition in Equation \eqref{eq:ri} gives
\begin{equation} \label{eq:ri2}
\begin{split}
    r_{i+1} = \argmin_{0\leq r\leq c\sqrt{n}}{\boldsymbol{y}^{\top}\boldsymbol{U}\boldsymbol{D}_{n}^2 (\boldsymbol{D}_{n} + r \mathds{I}_n)^{-2}\boldsymbol{U}^{\top}\boldsymbol{y} - 2\boldsymbol{\Tilde{y}}_{r_i}^{\top}\boldsymbol{U}\boldsymbol{D}_{n}(\boldsymbol{D}_{n} + r \mathds{I}_n)^{-1}\boldsymbol{U}^{\top}\boldsymbol{y}}.
\end{split}
\end{equation}

Equation \eqref{eq:ri2} can be solved by equating the derivative to $0$ as follows:
\begin{equation} \label{eq:ri3}
\begin{split}
-2\boldsymbol{y}^{\top}\boldsymbol{U}\boldsymbol{D}_{n}^2 (\boldsymbol{D}_{n} + r_{i+1} \mathds{I}_n)^{-3}\boldsymbol{U}^{\top}\boldsymbol{y} + 2\boldsymbol{\Tilde{y}}_{r_i}^{\top}\boldsymbol{U}\boldsymbol{D}_{n}(\boldsymbol{D}_{n} + r_{i+1} \mathds{I}_n)^{-2}\boldsymbol{U}^{\top}\boldsymbol{y} = 0.
\end{split}
\end{equation}
The roots of Equation \eqref{eq:ri3} may not have a simple closed form solution. Therefore, we use Newton's method to solve Equation \eqref{eq:ri3}.

Algorithm \ref{alg:1} below outlines and summarizes the full process of selecting the tuning parameter $r$ in the Air-HOLP method. Unless otherwise mentioned, the default inputs for Algorithm \ref{alg:1} that we use in our empirical research are $r_0 = 10, m' = \lceil n/\log(n) \rceil, c = 1000, \delta = 0.01,$ and maximum number of iterations $q_{max} = 10$.

\begin{algorithm}
\caption{Selection of the ridge tuning parameter $r$}\label{alg:1}
\begin{algorithmic}
\Require $\boldsymbol{X}$, $\boldsymbol{y}$, $r_0$, $m'$, c, $\delta, q_{max}$
\Ensure $\hat{r}$
\State Step 1: Initialize $i = 0 \rightarrow r_i = r_0$.
\State Step 2: Compute the Ridge-HOLP estimator $\boldsymbol{\hat{\beta}}_{r_i} = \boldsymbol{X}^\top(\boldsymbol{X}\boldsymbol{X}^\top + r_i \mathds{I}_n)^{-1}\boldsymbol{y}$.
\State Step 3: Rank the features in $\boldsymbol{X}$ by the absolute values of their estimated coefficients $|\boldsymbol{\hat{\beta}}_{r_i}|$. Then screen the top $m'$ features.
\State Step 4: Fit an ordinary least squares regression model utilizing the screened features, and compute the fitted response $\boldsymbol{\Tilde{y}}_{r_i}$.
\State Step 5: Solve Equation \eqref{eq:ri3} with Newton's method. If the solution is greater than $c \sqrt{n}$, set $r_{i+1} = c \sqrt{n}$.
\State Step 6: If $|r_{i+1} - r_{i}| < \delta r_{i+1}$ or if $i \geq q_{max}$, then output $r_{i+1}$. Otherwise, set $i = i + 1$ and repeat Steps 2 to 6.
\end{algorithmic}
\end{algorithm}

Once the tuning parameter is selected by Algorithm \ref{alg:1}, the Air-HOLP method computes the final Ridge-HOLP estimator $\boldsymbol{\hat{\beta}}_{\hat{r}} = \boldsymbol{X}^\top(\boldsymbol{X}\boldsymbol{X}^\top + \hat{r} \mathds{I}_n)^{-1}\boldsymbol{y}$, and ranks the features in $\boldsymbol{X}$ by the absolute values of their estimated coefficients $|\boldsymbol{\hat{\beta}}_{\hat{r}}|$. Then screen the top $m$ features. Note that $m$ does not need to be the same as $m'$ in Algorithm \ref{alg:1}.
Both $m$ and $m'$ are screening thresholds but they serve different purposes. The threshold $m'$ is used to efficiently estimate $\mathbf{y}_0$ which leads to a suitable and adaptive selection of the tuning parameter $r$. In contrast, the threshold $m$ is applied later with the goal of eliminating irrelevant features while retaining all true features. One distinction is that $m'$ is required to be less than $n$ while $m$ is not.\\

Although the focus in this paper is on Air-HOLP itself as a screening method, we expect that the implementation of Air-HOLP within group HOLP, Dynamic Tilted Current Correlation, and HOLP-DF will also lead to improvements.

\section{Simulation Study} \label{S4}

In this section, we empirically demonstrate the good performance of Air-HOLP. We primarily compare Air-HOLP's screening performance to Ridge-HOLP with fixed tuning parameter $r = 10$ as recommended by \cite{wang2016high}. While we initially also include Sure Independence Screening (SIS) in our comparison, our main focus is to showcase how Air-HOLP advances 
Ridge-HOLP by using an adaptive tuning parameter in the more general case when features are correlated. In all simulations, Air-HOLP is implemented with an initial tuning parameter $r_0 = 10$, selected tuning parameter $\hat{r} \in [0, 1000\sqrt{n}]$, and screening threshold $m = \lceil n/\log(n) \rceil$. The full simulation results and the code implementing Air-HOLP are both made available on GitHub at \url{https://github.com/Logic314/Air-HOLP.git}.

\subsection{Simulation Setup} \label{S4S1}

We generate $5,600$ unique $\boldsymbol{X}$ samples across $112$ distinct simulation settings, with $50$ samples per setting. For each distinct $\boldsymbol{X}$, we generate $250$ $\boldsymbol{y}$'s  across $25$ distinct simulation settings. This generates $1,400,000$ datasets [$\boldsymbol{y},\boldsymbol{X}$], encompassing $2,800$ distinct simulation settings, with $500$ samples per setting. The specific simulation settings employed are as follows.

We generate the design matrix $\boldsymbol{X}$ of size $n \times p$ from a multivariate normal distribution with covariance matrix $\boldsymbol{\Sigma} = (1-\rho)\mathds{I}_p + \rho$ for varying parameters $n$, $p$, and $\rho$.
\begin{itemize}
    \item Sample size $n$: $125, 250, 500, 1000$
    \item Number of features $p$: $250, 1250, 5000, 15000$
    \item Correlation between features $\rho$: $0, 0.3, 0.6, 0.9$
\end{itemize}

In addition to the compound symmetry correlation structure above, we also utilize a spatial correlation structure where only the middle $20 \%$ of the features are correlated while the remaining features are independent. Specifically, the covariance matrix of the middle $20 \%$ of the features is given by $\boldsymbol{\Sigma} = (1-\rho)\mathds{I}_{0.2p} + \rho$. The parameters $n$, $p$, and $\rho$ vary similarly in both cases of compound symmetry and spatial correlation.

We generate the response vector $\boldsymbol{y}$ by the linear regression model in \eqref{eq:linear} with varying parameters:
\begin{itemize}
    \item Number of true features $p_0$: $3, 6, 9, 12, 15$
    \item Theoretical $R^2$: $0.25, 0.5, 0.75, 0.9, 0.95$
\end{itemize}
In the case of compound symmetry, where all features have the same distribution, we assign the first $p_0$ features to be the true features. However, in the case of spatial correlation, we randomly assign the true features. The theoretical $R^2$ is defined as $R^2 = \operatorname{var}(\boldsymbol{x}^\top \boldsymbol{\beta})/\operatorname{var}(y)$ \citep{wang2016high}. The error term $\mathbfcal{E} \sim \operatorname{N}(0,\sigma^2\mathds{I}_n)$ where $\sigma^2$ is chosen to control the theoretical $R^2$, that is $\sigma^2 = (1 - R^2)/R^2 \times \operatorname{var}(\boldsymbol{x}^\top \boldsymbol{\beta})$.
Moreover, the non-zero regression coefficients vary randomly across the $\boldsymbol{y}$ samples and are generated by
$$\beta_j = (-1)^{u_j}(|z_j|+4\log(n)/\sqrt{n}),$$
where $z_j \sim \operatorname{Normal}(0,1)$ and $u_j \sim \operatorname{Bernoulli}(0.4)$ for $j = 1, \ldots, p_0$. This formula was introduced by \cite{fan2008sure} and used by \cite{fan2009ultrahigh} and \cite{wang2016high}. The $4\log(n)/\sqrt{n}$ term ensures sufficient signal to facilitate sure screening, while $|z_j|$ ensures sufficient variability between coefficients.

\subsection{Evaluation Metrics} \label{S4S2}
We compare the three methods Air-HOLP, Ridge-HOLP, and SIS using two measures of screening performance.
\begin{itemize}
    \item Sure Screening Threshold: The minimum model size needed to guarantee the inclusion of all true features. For instance, if $p_0 = 3$ and the rankings of the true features by the SIS method are $37$, $12$, and $54$. Then, the sure screening threshold of the SIS method is $54$, because we need to screen $54$ features to include all the $3$ true features. Mathematically, the sure screening threshold is given by $\max(s_j)$ for $j = 1, \ldots, p_0$, where $s_j$ is the ranking of the $j_{th}$ true feature by the screening method.
    \item Sure Screening Probability: The proportion of simulation runs where all true features are successfully included. Similarly to \cite{fan2008sure}, we screen the top $m = \lceil n/\log(n) \rceil$ features. Mathematically, the sure screening probability is given by
    $$\frac{1}{B}\sum_{b=1}^{B}{\mathds{1}(\max_j(s_{j,b}) \leq \lceil n/\log(n) \rceil)}, j = 1, \ldots, p_0,$$
    where $s_{j,b}$ is the ranking of the $j_{th}$ true feature in the $b_{th}$ simulation run, and $B = 500$ as there are $500$ sample pairs ($\boldsymbol{X}$, $\boldsymbol{y}$) per simulation setting.
\end{itemize}
We visualize the sure screening threshold using box plots and the sure screening probability using heat maps and line plots.

\subsection{Results} \label{S4S3}

Here we analyze the results from all $2,800$ simulation settings to assess the overall performance and comparison of the competing methods. Our analysis begins by demonstrating the effect of correlation on Air-HOLP, Ridge-HOLP, and SIS. Then, we focus on comparing Air-HOLP to Ridge-HOLP aiming to provide a fair and balanced narrative of the overall performance and comparison between them by showing results for representative settings that highlight differences but we conciously refrained from showing settings that give maximal difference between the methods.

\begin{figure}[htbp]
    \begin{subfigure}[b]{0.49\textwidth}
        \centering
        \textbf{Setting 1 - Compound symmetry}\par\medskip
        \includegraphics[width=\textwidth]{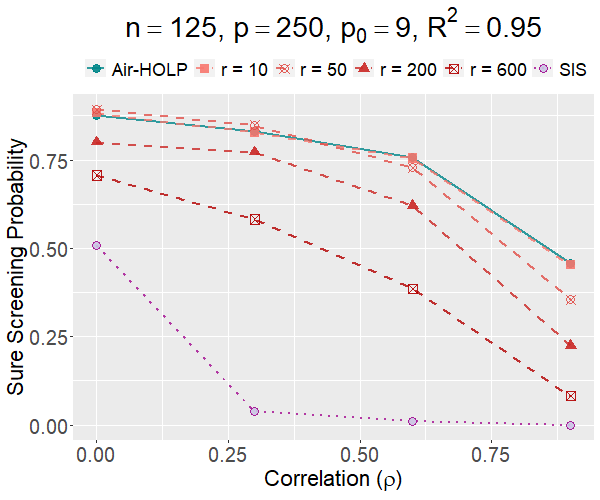}
        \caption{}
        \label{fig:4a}
    \end{subfigure}
    \hfill
    \begin{subfigure}[b]{0.49\textwidth}
        \centering
        \textbf{Setting 2 - Compound symmetry}\par\medskip
        \includegraphics[width=\textwidth]{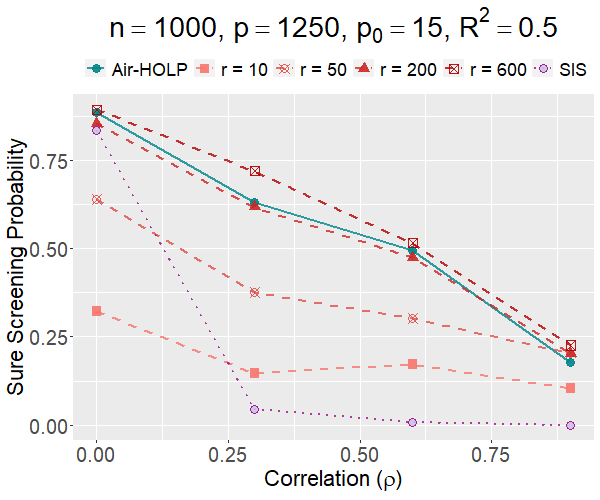}
        \caption{}
        \label{fig:4b}
    \end{subfigure}
    \hfill
    \begin{subfigure}[b]{0.49\textwidth}
        \centering
        \textbf{Setting 1 - Spatial correlation}\par\medskip
        \includegraphics[width=\textwidth]{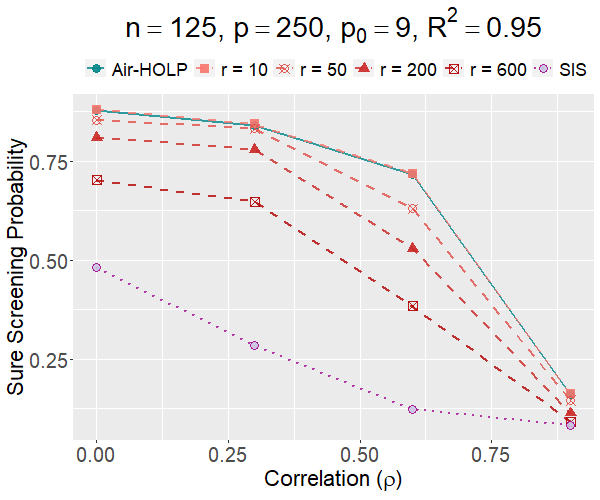}
        \caption{}
        \label{fig:4c}
    \end{subfigure}
    \hfill
    \begin{subfigure}[b]{0.49\textwidth}
        \centering
        \textbf{Setting 2 - Spatial correlation}\par\medskip
        \includegraphics[width=\textwidth]{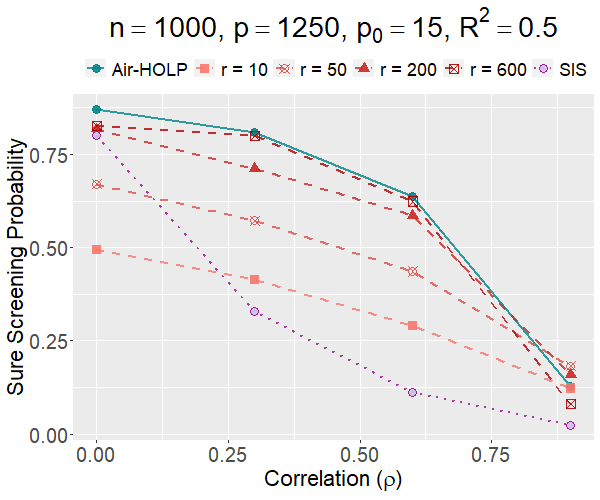}
        \caption{}
        \label{fig:4d}
    \end{subfigure}
    \caption{Line plots comparing the performances of Air-HOLP, Ridge-HOLP and SIS for different levels of correlation in four different simulation settings. For Ridge-HOLP, four values for the tuning parameter are compared: $r = 10, 50, 200, 600$.
    }
    \label{fig:4}
\end{figure}

The SIS by construction works best when features are uncorrelated but can have poor performance when features are highly correlated \citep{fan2008sure}. In contrast, Ridge-HOLP captures joint relationships between features and the response \citep{wang2016high} and is superior to SIS in highly correlated settings. We confirm this in Figure \ref{fig:4}, which also shows the strong performance of Air-HOLP when features are correlated.

In addition, Figure \ref{fig:4} also demonstrates the advantage of adaptive regularization by showcasing four settings were the optimal tuning parameter differ significantly. The smaller tuning parameters ($r = 10$ and $r = 50$) perform well in Figures \ref{fig:4a} and \ref{fig:4c} but not as well in Figures \ref{fig:4b} and \ref{fig:4d}, whereas the larger tuning parameters ($r = 200$ and $r = 600$) show the opposite trend. Air-HOLP, on the other hand, consistently performs well in all four settings. On the other hand, when comparing the compound symmetry settings (Figures \ref{fig:4a} and \ref{fig:4b}) to the spatial correlation settings (Figures \ref{fig:4c} and \ref{fig:4d}) we find that both Air-HOLP and Ridge-HOLP perform better when all features are highly correlated ($\rho = 0.9$) compared to when only the middle $20 \%$ are correlated. This suggests that the ridge penalty is more effective when the correlations between the features are at a similar level.

\begin{figure}[htbp]
    \includegraphics[width=\textwidth]{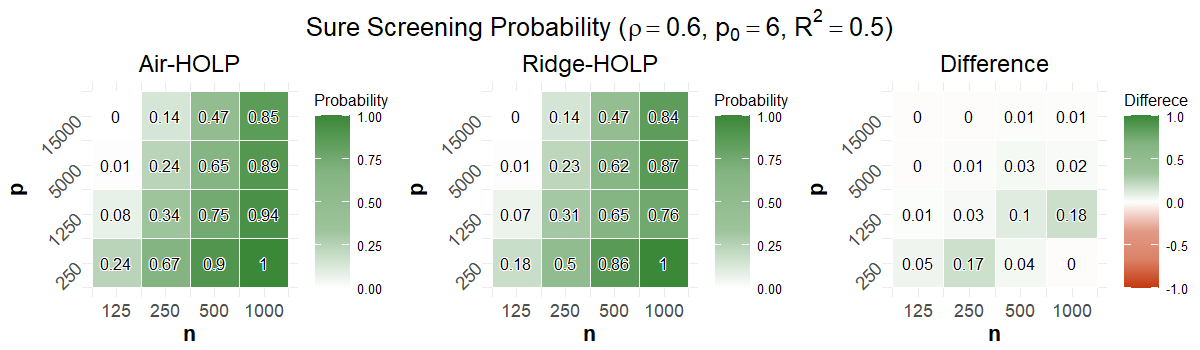}
    \caption{Sure Screening Probability Heatmaps. The left panel shows Air-HOLP, the middle panel shows Ridge-HOLP (with $r = 10$), and the right panel shows the difference between Air-HOLP and Ridge-HOLP (Sure Screening Probability of Air-HOLP minus Ridge-HOLP). Results shown are for the compound symmetry correlation settings.}
    \label{fig:5}
\end{figure}

We now shift the focus to comparing Air-HOLP and Ridge-HOLP with the recommended tuning parameter of $r = 10$, as suggested by \cite{wang2016high}. We mainly focus on the compound symmetry simulation results. However, the same conclusions and insights we discuss apply to the spatial correlation case.
Across all $1600$ compound symmetry simulation settings, the difference in sure screening probability between Air-HOLP and Ridge-HOLP ranged from $-0.02$ to $0.60$. In $47\%$ of the settings, the sure screening probabilities for Air-HOLP and Ridge-HOLP were equal, typically occurring when both probabilities were either $0$ or $1$. In $45\%$ of the settings, Air-HOLP's sure screening probability was higher than Ridge-HOLP's, while in $8\%$ of the settings, it was lower. To better understand the factors influencing this gap in performance, we analyze the impact of $n, p, p_0$, and $R^2$ on feature screening performance of Air-HOLP and Ridge-HOLP.

Figure \ref{fig:5} showcases the joint effect of the sample size $n$ and the number of features $p$ on the screening performance. The settings $\rho = 0.6, p_0 = 6, R^2 = 0.5$ for the presented heat maps in Figure \ref{fig:5} were selected to ensure a diverse range of sure screening probabilities from $0$ to $1$ across the values of $n$ and $p$ for both Air-HOLP and Ridge-HOLP. The selected settings provide a holistic view of the joint impact of $n$ and $p$ on screening performance and comparison. Figure \ref{fig:5} reveals that the performance gap between Air-HOLP and Ridge-HOLP is more pronounced when $n$ is close to $p$ (e.g., when $n = 1000$ and $p = 1250$). However, the difference can still be significant in other settings (e.g., when $n = 500$ and $p = 1250$). Moreover, Figure \ref{fig:5} shows that for both Air-HOLP and Ridge-HOLP, a large $n$ has a strong positive impact on screening performance, while a large $p$ has a smaller negative influence. This demonstrates the strong capability of both methods to handle high-dimensional data.

\begin{figure}[htbp]
    \begin{subfigure}[b]{0.49\textwidth}
        \includegraphics[width=\textwidth]{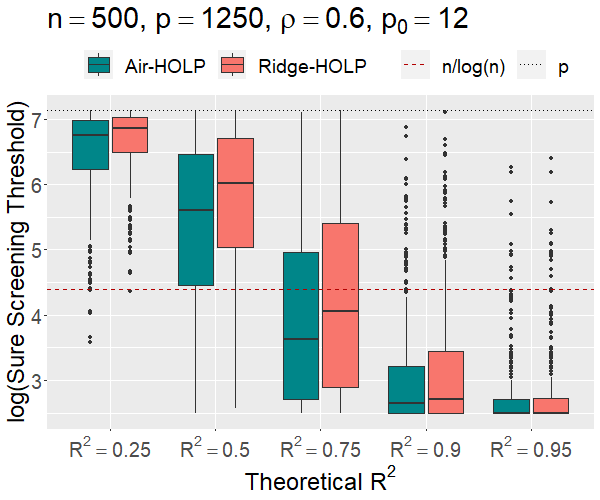}
        \caption{The logarithm of the Sure Screening Threshold vs. Theoretical $R^2$}
        \label{fig:6a}
    \end{subfigure}
    \hfill
    \begin{subfigure}[b]{0.49\textwidth}
        \includegraphics[width=\textwidth]{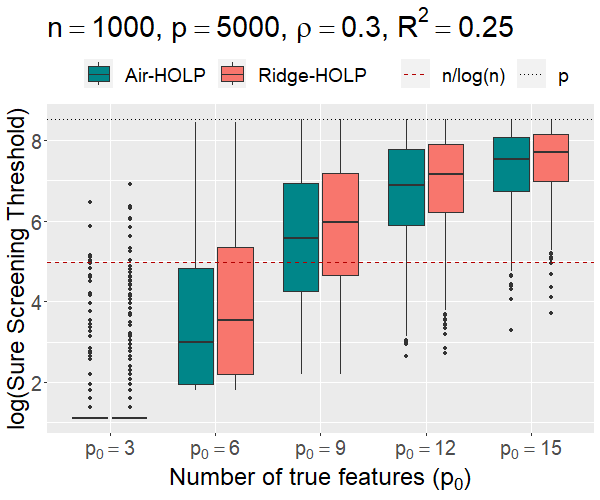}
        \caption{The logarithm of the Sure Screening Threshold vs. the number of true features $p_0$}
        \label{fig:6b}
    \end{subfigure}
    \caption{Boxplots comparing Air-HOLP to Ridge-HOLP (with $r = 10$) for different values of $R^2$ and $p_0$. The dashed line represents the screening threshold $\lceil n/\log(n) \rceil$. Values below this line indicate simulation runs where all true features are screened in. Results shown are for the compound symmetry correlation settings.}
    \label{fig:6}
\end{figure}

Figure \ref{fig:6} showcases the effect of both the number of true features $p_0$ and the theoretical $R^2$ on the screening performance. The settings $n = 500, p = 1250, \rho = 0.6, p_0 = 12$ for Figure \ref{fig:6a} and $n = 1000, p = 5000, \rho = 0.3, R^2 = 0.25$ for Figure \ref{fig:6b} were selected to ensure a diverse range of sure screening thresholds across the values of $p_0$ and $R^2$ for both Air-HOLP and Ridge-HOLP. These settings provide a comprehensive view of the impact of $p_0$ and $R^2$ on screening performance. Figure \ref{fig:6b} illustrates that the performance of both Air-HOLP and Ridge-HOLP declines as the number of true features $p_0$ increases. Conversely, Figure \ref{fig:6a} suggests a direct relationship between the theoretical $R^2$, and the screening performance of both methods. Furthermore, Air-HOLP consistently outperforms Ridge-HOLP across most settings of $p_0$ and $R^2$.

In conclusion, both Air-HOLP and Ridge-HOLP outperform SIS in correlated settings. However, Air-HOLP demonstrates a consistent advantage over Ridge-HOLP across various settings of $n$, $p$, $\rho$, $p_0$, and $R^2$, achieving higher sure screening probabilities and lower sure screening thresholds as demonstrated in figures \ref{fig:4}-\ref{fig:6}. This showcases the effectiveness of Air-HOLP's adaptive tuning parameter selection in enhancing feature screening performance.

\section{Application to Prostate Cancer Data} \label{S5}

We apply Air-HOLP, Ridge-HOLP and SIS to the \href{https://schlieplab.org/Static/Supplements/CompCancer/CDNA/tomlins-2006-v2/index.html}{Tomlins-V2} prostate cancer genetic data of \cite{tomlins2006integrative} consisting of $n = 92$ samples and $p = 1288$ genes. The objective of this dataset is to understand the genetic changes that occur as prostate cancer progresses through different stages. The dataset consists of genetic information collected from four stages of prostate cancer.

\begin{itemize}
    \item Benign Epithelium: Normal, healthy prostate cells.
    \item Prostatic Intraepithelial Neoplasia: Early changes in cells that might become cancer.
    \item Prostate Cancer: Actual cancer cells.
    \item {Metastatic Disease:} Cancer cells that have spread to other body parts.
\end{itemize}

Our objective is to compare how well the feature screening methods capture the joint relations between the genes and the responses. We first screen $\lceil n/\log(n) \rceil = 21$ genes for each of the four binary responses by each of the three methods. Then, we measure the joint relations between the screened features and the response through the coefficient of multiple correlation, given by 
$$\text{Multiple R}  = \sqrt{\frac{\|\boldsymbol{\hat{y}} - \boldsymbol{\bar{y}} \|_2^2}{\|\boldsymbol{y} - \boldsymbol{\bar{y}} \|_2^2}},$$
where $\boldsymbol{\bar{y}}$ is the mean of $\boldsymbol{y}$ and $\boldsymbol{\hat{y}}$ is the fitted response for a given set of screened features. 

\begin{figure}[htbp]
    \begin{subfigure}[b]{0.49\textwidth}
        \includegraphics[width=\textwidth]{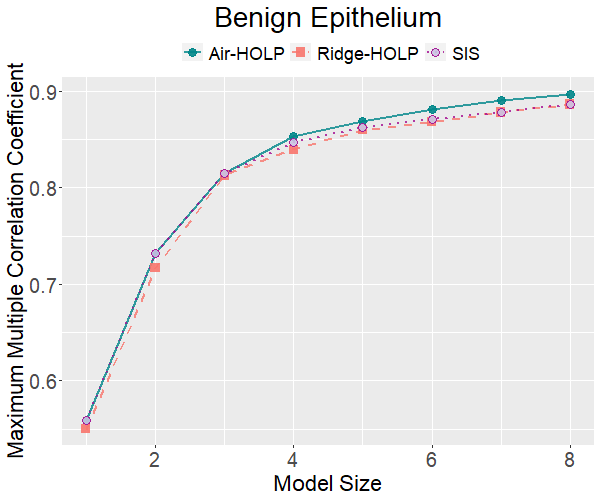}
        \caption{}
        \label{fig:7a}
    \end{subfigure}
    \hfill
    \begin{subfigure}[b]{0.49\textwidth}
        \includegraphics[width=\textwidth]{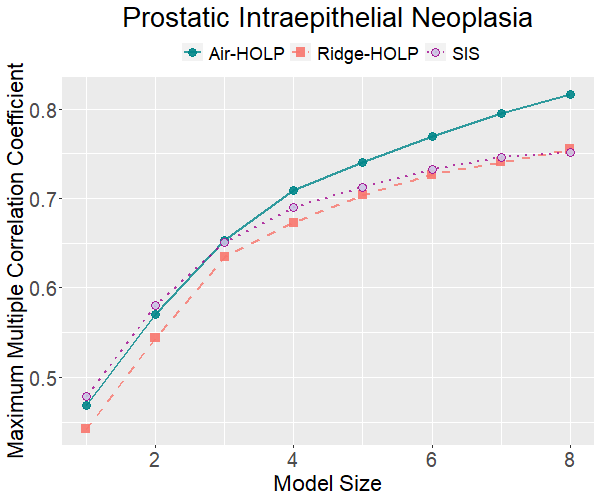}
        \caption{}
        \label{fig:7b}
    \end{subfigure}
    \begin{subfigure}[b]{0.49\textwidth}
        \includegraphics[width=\textwidth]{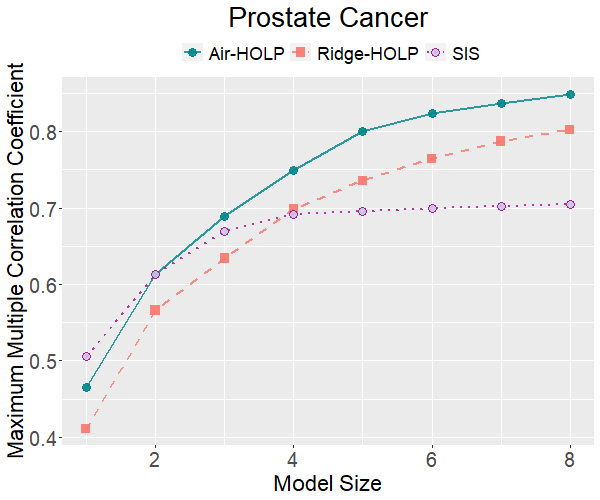}
        \caption{}
        \label{fig:7c}
    \end{subfigure}
    \hfill
    \begin{subfigure}[b]{0.49\textwidth}
        \includegraphics[width=\textwidth]{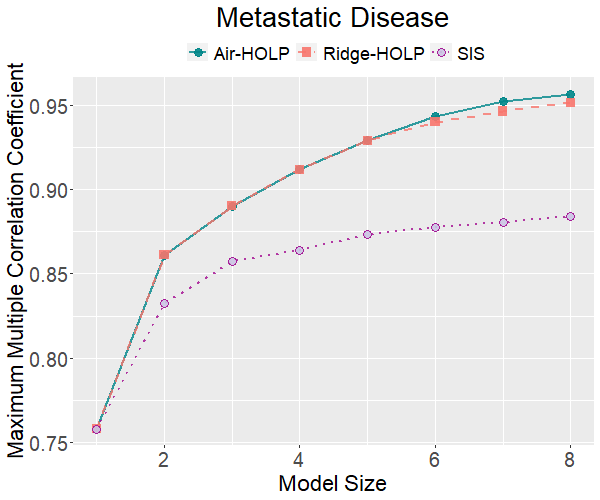}
        \caption{}
        \label{fig:7d}
    \end{subfigure}
    \caption{Maximum Multiple R vs. model size using the screened genes by the competing methods Air-HOLP, Ridge-HOLP, and SIS}
    \label{fig:7}
\end{figure}

The largest Multiple R for model size $k=1,\ldots,8$ is visualised in Figure \ref{fig:7}. This figure shows that Ridge-HOLP and Air-HOLP outperform SIS, because features are correlated in this dataset for all four stages of prostate cancer progression. The maximum Multiple R for Ridge-HOLP and Air-HOLP is higher than that of SIS, especially for larger models. We also observe that for this dataset, Air-HOLP consistently outperforms Ridge-HOLP.

\section{Computational Complexity} \label{S6}

Air-HOLP shares the same time complexity as Ridge-HOLP. However, Air-HOLP employs an iterative process to update the initial tuning parameter. We investigate how this iterative process impacts the speed of Air-HOLP. 

\begin{figure}[h]
    \centering
    \begin{subfigure}{0.49\textwidth}
    \includegraphics[width=0.995\linewidth]{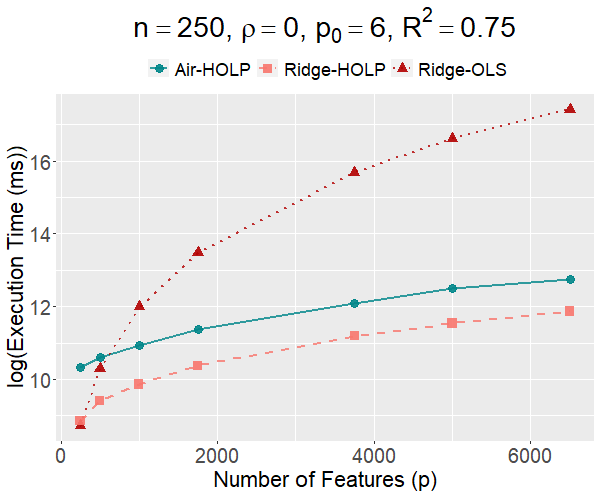} 
    \caption{The logarithm of execution time in milliseconds vs. number of features $p$}
    \label{fig:2.1}
    \end{subfigure}
    \begin{subfigure}{0.49\textwidth}
    \includegraphics[width=0.995\linewidth]{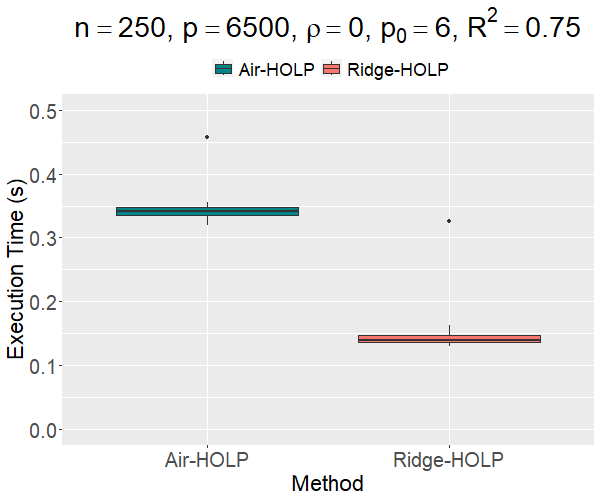} 
    \caption{Execution time of Air-HOLP vs. Ridge-HOLP in seconds}
    \label{fig:2.2}
    \end{subfigure}
    \caption{Compression of the execution time between Ridge-OLS, Ridge-HOLP, and Air-HOLP. On the left, we present the trimmed mean of the logarithm of execution time in milliseconds for $10$ simulations ($10 \%$ trimming). On the right, we present a boxplot of the execution time in seconds for $10$ simulations}
    \label{fig:2}
\end{figure}

Notably, the eigen decomposition of $\boldsymbol{X}\boldsymbol{X}^\top$ in Equation \eqref{eq:eigen} is the only step in Air-HOLP with complexity $O(n^2p + n^3)$, and this step is not repeated. The remaining operations in Air-HOLP have a complexity of $O(np + n^2)$ or less. To empirically examine the speed and time complexity of Air-HOLP, we compare its execution time to that of Ridge-HOLP and ordinary ridge regression (Ridge-OLS). We generate samples of $\boldsymbol{X}$ and $\boldsymbol{y}$ similarly as in Section \ref{S4S1} and use a $12_{th}$ Gen Intel Core i7-1255U ($2$ P-cores, $8$ E-cores, $12$ threads, $1.70$ GHz base frequency, up to $4.70$ GHz turbo frequency). Figure \ref{fig:2.1} confirms that Air-HOLP has the same time complexity as Ridge-HOLP and both methods are considerably faster than Ridge-OLS in high-dimensional settings where $p >> n$. Figure \ref{fig:2.2} shows that Air-HOLP takes roughly twice as long as Ridge-HOLP.

\section{Discussion} \label{S7}

Traditional feature screening methods such as SIS measure the marginal relations between features and the response, which can fail when features are correlated. Ridge-HOLP was developed to address this issue and we have now further built on Ridge-HOLP by the development of Air-HOLP, which incorporates adaptively selecting the ridge tuning parameter. This adaptive selection aims to minimize prediction error, thereby effectively balancing the bias-variance trade-off and enhancing feature screening performance.

Through extensive simulation studies, we demonstrated that Air-HOLP consistently outperforms Ridge-HOLP with a fixed tuning parameter across various simulation settings. Air-HOLP achieves higher sure screening probabilities and lower sure screening thresholds. The advantage of Air-HOLP is more pronounced when the number of samples $n$ is closer to the number of features $p$. Additionally, both Air-HOLP and Ridge-HOLP outperformed SIS in correlated settings, making Air-HOLP especially useful for identifying relevant features in high-dimensional correlated data. However, the ridge penalty is more effective when the correlations between the features are at a similar level. Moreover, Air-HOLP has the same time complexity as Ridge-HOLP and also enjoys being computationally efficient. We also found that in most simulations, Air-HOLP converged as quickly as in $3$ to $6$ iterations. Therefore, we set the maximum number of iterations to be $10$ in our code. We further applied Air-HOLP, Ridge-HOLP and SIS to a prostate cancer genetic dataset. The results confirmed the good performance of Air-HOLP.

While Air-HOLP provides significant advantages for feature screening, certain limitations should be noted. Both Air-HOLP and Ridge-HOLP are designed to measure linear relationships between features and the response and may fail to capture non-linear dependencies. Additionally, these methods are optimized for cases where $p > n$. Although they are functional when $n > p$, their computational efficiency is not ideally suited for such scenarios. These limitations suggest potential areas for future improvement. Moreover, the iterative framework of Algorithm \ref{alg:1}, while specifically designed for the ridge penalty, may be extended and adapted to other L2-regularized penalties. Future research could explore these extensions as well as generalise Air-HOLP to non-linear models.

\section*{Acknowledgments}
Samuel Muller was supported by the Australian Research Council Discovery Project Grant (DP230101908).

\section*{Supplementary information}
The full simulation results and the code implementing Air-HOLP are made available on GitHub at \url{https://github.com/Logic314/Air-HOLP.git}.

\section*{Author contributions}
Ibrahim Joudah conceived the original research idea, developed the R codes, and conducted the simulation study. Samuel Muller and Houying Zhu provided supervision and guidance throughout the project. All authors collaboratively contributed to developing the research methodology, including refining the Air-HOLP method, designing the simulation study, and applying to real-world data. All authors contributed to writing and revising the manuscript.

\bibliography{Air-HOLP}
\end{document}